\documentclass[showpacs,aps,pra,floatfix,reprint,superscriptaddress,footinbib,citeautoscript]{revtex4-2}
\usepackage{amssymb,amsfonts,amsmath}
\usepackage{graphicx}
\usepackage{verbatim}
\usepackage[utf8]{inputenc}
\usepackage[T1]{fontenc}
\usepackage{soul,xcolor,colortbl}
\usepackage{multirow}
\usepackage{bm}
\usepackage{array}
\usepackage{chemmacros}
\usepackage{color}
\usepackage{hyperref} \hypersetup{colorlinks=true,citecolor=blue,linkcolor=blue,urlcolor=blue}
\usepackage{mdframed}
\usepackage{longtable}
\usepackage{enumitem}
\usepackage{booktabs}
\usepackage{lipsum}
\usepackage{orcidlink}

\newcolumntype{L}[1]{>{\raggedright\arraybackslash}p{#1} }
\newcolumntype{C}[1]{>{\centering \arraybackslash}p{#1} }
\newcolumntype{R}[1]{>{\raggedleft \arraybackslash}p{#1} }

\DeclareSIUnit\inch{inches}

\def\AFLOW{{\small AFLOW}}
\def\AFLOWorg{{\tt\small aflow.org}}
\def\SISSO{{\small SISSO}}
\def\BMG{{\small BMG}}
\def\BMGs{{\small BMG}s}

\setcitestyle{square}

\makeatletter \renewcommand\frontmatter@abstractwidth{\dimexpr\textwidth\relax} \makeatother

\def\MEMS{Department of Mechanical Engineering and Materials Science, Duke University, Durham, NC 27708, USA}
\def\CEM{Center for Extreme Materials, Duke University, Durham, NC 27708, USA}
\def\YALE{Department of Mechanical Engineering and Materials Science, Yale University, New Haven, CT 06511, USA}
\def\MSEW{Department of Materials Science and Engineering, University of Wisconsin - Madison, Madison, WI 53706, USA}
\def\BROWN{School of Engineering, Brown University, Providence, RI 02912, USA}

\begin{document}
\title{Soliquidy: {a descriptor for atomic geometrical confusion}}
\author{Hagen~Eckert\,\orcidlink{0000-0003-4771-1435}}\affiliation{\MEMS}\affiliation{\CEM}
\author{Sebastian~A.~Kube\,\orcidlink{0000-0001-8167-4178}}\affiliation{\YALE}\affiliation{\MSEW}
\author{Simon~Divilov\,\orcidlink{0000-0002-4185-6150}}\affiliation{\MEMS}\affiliation{\CEM}
\author{Asa~Guest}\affiliation{\MEMS}\affiliation{\CEM}
\author{Adam~C.~Zettel\,\orcidlink{0000-0003-1645-9476}}\affiliation{\MEMS}\affiliation{\CEM}
\author{David~Hicks\,\orcidlink{0000-0001-5813-6785}}\affiliation{\MEMS}\affiliation{\CEM}
\author{Sean~D.~Griesemer\,\orcidlink{0000-0001-5531-0725}}\affiliation{\MEMS}\affiliation{\CEM}
\author{Nico~Hotz\,\orcidlink{0009-0008-2469-2693}}\affiliation{\MEMS}\affiliation{\CEM}
\author{Xiomara~Campilongo\,\orcidlink{0000-0001-6123-8117}}\affiliation{\CEM}
\author{Siya~Zhu\,\orcidlink{0000-0003-0394-4050}}\affiliation{\BROWN}
\author{Axel~van de Walle\,\orcidlink{0000-0001-5877-6782}}\affiliation{\BROWN}
\author{Jan~Schroers\,\orcidlink{0000-0001-5877-6782}}\affiliation{\YALE}
\author{Stefano~Curtarolo\,\orcidlink{0000-0003-0570-8238}}\email[]{stefano@duke.edu}\affiliation{\MEMS}\affiliation{\CEM}

\date{\today}

\begin{abstract}
 \noindent
 Tailoring material properties often requires understanding the solidification process.
 Herein, we introduce the geometric descriptor Soliquidy, which numerically captures the Euclidean transport cost between the translationally disordered versus ordered {states of a} materials.
 {As a testbed, we apply Soliquidy to the classification of glass-forming metal alloys.
 By extending and combining an experimental library of metallic thin-films
 (glass/no-glass)
 with the \AFLOWorg\ computational database (geometrical and energetic information of mixtures)
 we found that the combination of Soliquity and formation enthalpies generates an effective classifier for glass formation.
Such classifier is then used to tackle a public dataset of metallic glasses showing that the glass-agnostic
assumptions of Soliquity can be useful for understanding kinetically-controlled phase transitions.}
\end{abstract}
\maketitle

\noindent

\section*{Introduction}
\label{sec:intro}
Rational control of the properties of materials requires understanding the phase transitions involved in the synthesis of the material~\cite{yeomansStatisticalMechanicsPhase1992,jacksonKineticProcessesCrystal2004,fultzPhaseTransitionsMaterials2020}.
During the transition from liquid/gas/plasma to solid/crystalline, the movement of the atoms (transport) and the specific heat rate change (latent heat) determine the final outcome.
Balancing the interplay between these factors can lead to the formation of non-ground-states configurations, often technologically advantageous.
Despite the critical importance of kinetics, searches for new materials often focus primarily on the energetic landscape. Even theories of synthesizability are mostly discussed in terms of initial and final energetic states, neglecting the process used for the synthesis~\cite{deed}.
Until now, geometrical descriptors mostly aim to capture the interconnection of the atoms, or encode the crystalline structure of a material itself, as it is commonly used in machine learning approaches \cite{Drautz_ACE_PRB_2019,liEncodingAtomicStructure2022,chenGraphNetworksUniversal2019,russoGlassFormingAbility2018,curtarolo:art174,curtarolo:art128}.
{Hence, the need for characterizing atomic movements in liquid-to-solid transitions remains.}

In this article, we introduce Soliquidy ($\mathfrak{S}$), a geometrical descriptor capturing the optimal-transport aspect of the transition between translationally disordered (melt, plasma, solution, vapor - atom position is not fixed) versus ordered materials states in a single numerical value, measuring the {integrated Euclidean distance} between a structured and {an uniformly} randomized state.
{We test Soliquidy on bulk metallic glasses (\BMGs)~\cite{hofmannBulkMetallicGlasses2013, kruzicBulkMetallicGlasses2016, greerMetallicGlasses2023,telfordCaseBulkMetallic2004, schroersProcessingBulkMetallic2010, inoueRecentDevelopmentApplication2011,gaoRecentDevelopmentApplication2022}, perfect candidates since the rapid solidification needed for their formation emphasizes the kinetic (Greer's) ``confusion'' of atomic rearrangements~\cite{greer1993confusion}.}
{
During the first stage of solidification many different crystal structures are created concurrently: the more distinct, low-energy structures can energetically coexist, the more the alloy is likely to form an amorphous phase~\cite{curtarolo:art112,curtarolo:art154}.
This correlates with Soliquidy: geometrical organizations capable of moving atoms around equilibrium positions with small transport costs will be capable of generating many dissimilar structures with similar energy.}

{The approach is fourfold.
  {\bf i.}~We extend our experimental thin-film library of glasses~\cite{ding2014combinatorial, kubeMetastabilityHighEntropy2020, liDatadrivenDiscoveryUniversal2022, kubeCompositionalDependenceFragility2022}, analyze the mixtures and label the compositions as glass or ordered;
  {\bf ii.}~We equip such data with geometrical/energetic information extracted from the  the \AFLOWorg\ database~\cite{curtarolo:art190};
  {\bf iii.}~By using a machine learning/artificial intelligence approach, specifically the \underline{s}ure \underline{i}ndependence \underline{s}creening and \underline{s}parsifying \underline{o}perator (\SISSO)~\cite{SISSO}, we find several effective classifiers based on Soliquidy and enthalpy values (note that to account for the composition spread in the nucleation phase, we apply a new weighting scheme that treats this problem statistically).
  {\bf iv.}~We apply the {best found} {classifier} to an independently published dataset of \BMGs~\cite{wardMachineLearningApproach2018,wardBulkMetallicGlasses2018}.
  The final results demonstrate the usefulness of Soliquidy to tackle amorphous systems.
}

\begin{figure*}[ht!]
  \includegraphics[width=0.9\textwidth]{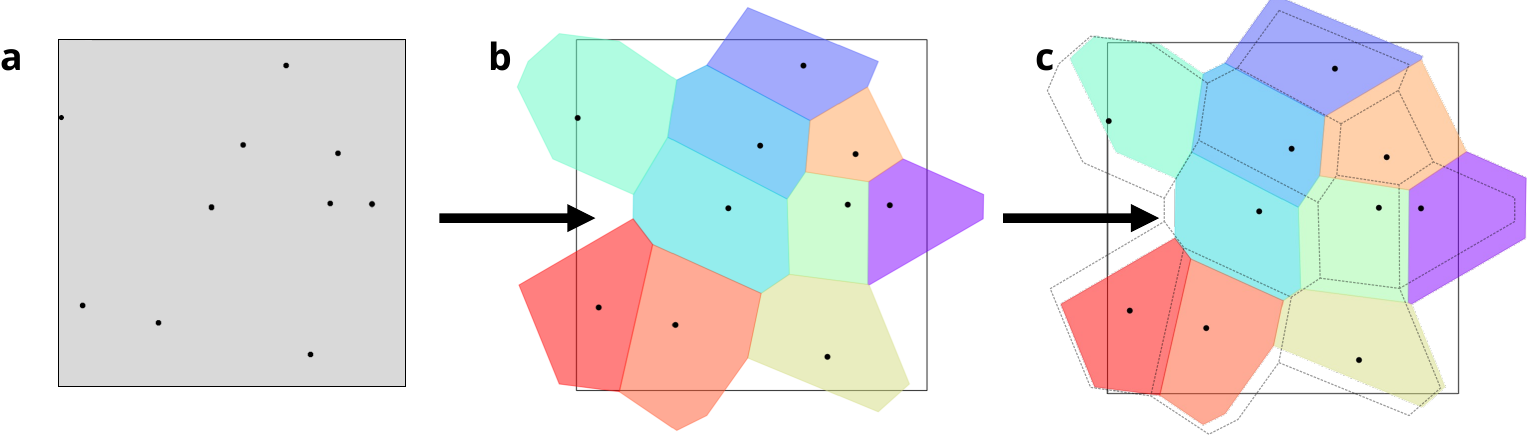}
  \caption{\small \textbf{Optimal transport cell generation}
    (\textbf{a}) the grey filling of the cell represents the translationally disordered state, while the black dots represent the center of atoms fixed in the solid state (randomly placed to emphasize different cell forms);
    (\textbf{b}) each atom is assigned a cell using Voronoi tessellation; and
    (\textbf{c}) final generated Voronoi cells with optimized growth rate resulting in equal portions of the cell area, the original cells are outlined with dashed gray lines.
  }
  \label{fig:con2dis}
\end{figure*}

\section*{Discussion}
\label{sec:theory}
\noindent {\bf Soliquidy $\mathfrak{S}$ } \\
The goal is to develop a robust geometric descriptor that captures the transition from a translationally disordered to an ordered (solid, crystalline - atom positions are fixed) state.
The descriptor should be based on the full atomic distribution in three dimensions without relying on symmetry, making it suitable for {even} structurally disordered systems.
The term Soliquidy derives from the merge of \underline{sol}id- and \underline{liquid}-like states.

We describe the transition as an optimal-transport problem~\cite{levyNotionsOptimalTransport2018, peyreComputationalOptimalTransport2019}, where the goal is to address the logistical problem of moving atoms at the minimal cost, functional of the distances between the atoms' positions from the translationally disordered state to the final ordered one.
{The idea is the following:
{\bf i.}~one sits on an atom and creates a Voronoi cell around it;
{\bf ii.}~the atom is then uniformly smeared inside such cell;
{\bf iii.}~an integral transport cost is defined for bringing the smeared atom back to its original position, where the cost is a functional of the traveled distance; and
{\bf iv.}~the procedure is repeated and summed for all the available atoms to give a macroscopic transport cost; i.e., Soliquity.}

\noindent{{\bf Voronoi tesselation } \\
The choice of Voronoi tesselation~\cite{voronoiNouvellesApplicationsParametres1908, wignerConstitutionMetallicSodium1933} comes from its uniquitous use in representing geometric properties of materials, for example:}
to estimate coordination numbers \cite{fukunagaVoronoiAnalysisStructure2006},
to capture structural heterogeneity \cite{wardIncludingCrystalStructure2017},
to define degenerate Delaunay clusters to study crystallization \cite{brostowVoronoiPolyhedraDelaunay1998},
and to generate representation of crystalline compounds in machine learning models \cite{blatov*VoronoiDirichletPolyhedra2004}.
However, as the {atoms of a specific element are distributed evenly over the observed volume} in the translationally disordered state, each atom needs to be assigned a surrounding cell with the same volume fraction {$V_{\mathrm{Vor}}=V/N$\,,
where $V$ and $N$ are the volume and the number of atoms of the cell.}
The standard Voronoi tessellation breaks this requirement.
This limitation is overcome by weighting the growth rate of the different Voronoi cells so that the resulting cells are all equal in size, as shown in the example in Figure~\ref{fig:con2dis}.
{For this purpose, we employ} the Kitagawa, M\`erigot and Thibert ({\small KMT}) algorithm~\cite{kitagawaConvergenceNewtonAlgorithm2019}, giving the unique distribution of weights assigned of the Voronoi cells.
This approach has no limitations when the chosen volume is periodic.
Otherwise, the shape and weight associated with Voronoi cells bordering the surface will depend on the chosen boundary --- yet the influence of the errors caused by the outer layer decreases with the number of {Voronoi} cells.
As such, large amorphous systems can still be tackled.
Nevertheless, all systems in this article are periodic, so the limitation does not appear.

\begin{figure}[ht!]
  \includegraphics[width=0.5\textwidth]{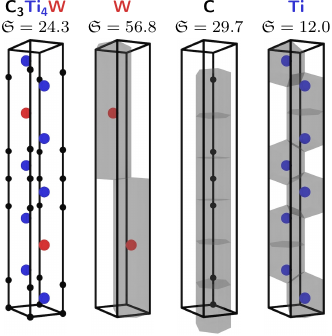}
  \caption{\small \textbf{Application of Soliquidy ($\mathfrak{S}$)} to \ch{C3Ti4W}
    (\href{https://aflow.org/material/?id=03cb24fa8c754e9d}{aflow:03cb24fa8c754e9d})
    in the spacegroup $P6_3/mmc$ (\#194). The volumes highlighted in
    gray for each element are the equal sized Voronoi cells, that form
    the basis of the Soliquidy descriptor.}
  \label{fig:soliquidy3D}
\end{figure}

\begin{figure}[ht!]
  \includegraphics[width=0.5\textwidth]{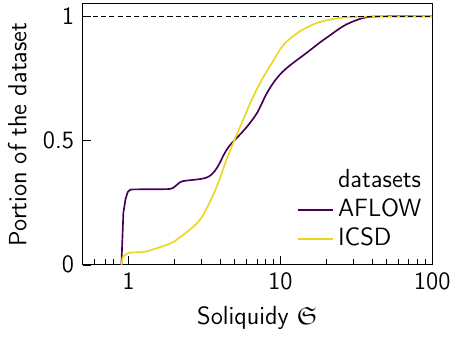}
  \caption{\small \textbf{Distribution of Soliquidy ($\mathfrak{S}$) values} in the whole \AFLOWorg{} database and a sub collection based the experimentally observed materials listed in the Inorganic Crystal Structure Database (ICSD)~\cite{ICSD, ICSD1}.}
  \label{fig:soliquidyDist}
\end{figure}

\noindent{{\bf Euclidean trasport cost } \\
Given the whole cell weigths/tesselation, the value of Soliquity for each atom can be calculated.
Integrating through the Voronoi volume $V_\mathrm{Vor}(i)$, around the $i$-atom and located in position $\mathbf{r}_i$, the transport cost $C(i)$ is defined from the distance of the smeared atom to the equilibrium position:
\begin{align}
  C(i) &\equiv \int_{V_\mathrm{Vor}(i)}
  \!\!\!\!\!\!\!\!\!\!\!\!\!\!
  |\mathbf{r}-\mathbf{r}_i|\,\mathrm{d^3}\mathbf{r}\,.
\end{align}
In such definition, we have chosen to associate the cost to the Euclidean distance --- linked to the typical transport costs; e.g., total fuel or time spent in moving an object \cite{peyreComputationalOptimalTransport2019}.
A normalization is necessary for comparisons based on shape. Thus, $C(i)$ is normalized by a sphere having a four-dimensional volume given by the three-dimensional {\it spherical radius} of the Voronoi cell. One gets $r_{\mathrm{S}}(i) \equiv \sqrt[3]{{3V_\mathrm{Vor}(i)}/{4\pi}}$ and $C_{\textrm{S}}(i) \equiv \pi r_{\mathrm{S}}^4$.}
{
  Soliquidy is then expressed as the sum over all atoms of the root mean squared difference of all normalized costs:
  \begin{equation}
    \mathfrak{S} \equiv 100 \cdot \sqrt{\sum_{i\in\mathrm{atoms}}\!\!\!\!{\left[C(i)/C_\mathrm{S}(i)-1\right]^2}/N}.
  \label{macrosoliquity}
\end{equation}
For multi-component systems, one can obtain a species-restricted Soliquity, by limiting the sum in Equation (\ref{macrosoliquity}) to atoms belonging to a given $j$-species, thus obtaining $\mathfrak{S}_j$. A concentration-average can also be performed to obtain the macroscopic Soliquity value:
\begin{equation}
  \mathfrak{S} \equiv\!\!\!\!\sum_{j\in\mathrm{species}}\!\!\!\!{x_j \mathfrak{S}_j}.
\end{equation}
An example of Soliquidy from the different species in \ch{C3Ti4W} is shown in Figure~\ref{fig:soliquidy3D}.}
There, $\mathfrak{S}_{\rm Ti}$ is significantly lower than $\mathfrak{S}_{\rm W}$.
This difference arises by the shape of the generated cells: close to spherical for \ch{Ti}, and elongated for \ch{W}.
The location of the atoms in their cells also affects the total $\mathfrak{S}$: the higher, the further from the center.

Figure~\ref{fig:soliquidyDist} depicts the distribution over the whole \AFLOWorg\ computational database.
The value of Soliquidy ranges from approximately \numrange{0.9}{100}.
The experimentally observed materials listed in the Inorganic Crystal Structure Database (ICSD) \cite{ICSD, ICSD1} show a smooth distribution with most materials having $\mathfrak{S}<10$.
When looking at \AFLOWorg\ dataset, including many computationally derived structures, the fraction of systems having $\mathfrak{S}<2$ is significant higher. The reason lies in historical convenience: i.e.,\ easy-to-calculate systems with few atoms per cell plus their combinatorial chemical combinations were the first to be added during the fast paced growth of the repository \cite{curtarolo:art173,eckertAFLOWLibraryCrystallographic2024,curtarolo:art135,curtarolo:art53}.
Conversely, the region with high Soliquidy is populated with more ``computationally-demanding'' complex structures (e.g.,\ like layered composites, multicomponent and disordered ceramics), and therefore they are appearing at lower rate \cite{deed,curtarolo:art201,curtarolo:art170,curtarolo:art187}.

\noindent {\bf Nuclei composition kernel } \\
As discussed above, entries in materials repository like \AFLOWorg{} are clustered at a limited number of compositions.
In contrast, the composition of a material can be finely controlled in an experimental setting.
Bridging this difference is crucial to using the enormous amount of data available in computational material databases.
Our approach {averages} entries in the database into ``virtual entries'' at any given concentration, to make predictions possible.
For this task, we introduce the nuclei composition kernel (NCK) to weigh database entries {according to their distance from the target concentration}.

In our chosen testbed of metallic glasses, the transition from the translationally disordered into the ordered solid phase with fixed atom positions occurs under a time constraint.
Numerous nucleation seeds with different composition and crystal systems are competing at the same time, and due to the high cooling rates, thermodynamic balancing does not have enough time.
This results in geometrical confusion, as proposed by Greer in 1993~\cite{greer1993confusion}.
If a high enough number of different phases with incompatible structures are created alongside each other, the growth from the nucleation sites is severely hindered, and the resulting material will lack long-range ordering.

\begin{figure}[ht!]
 \includegraphics[width=1.0\linewidth]{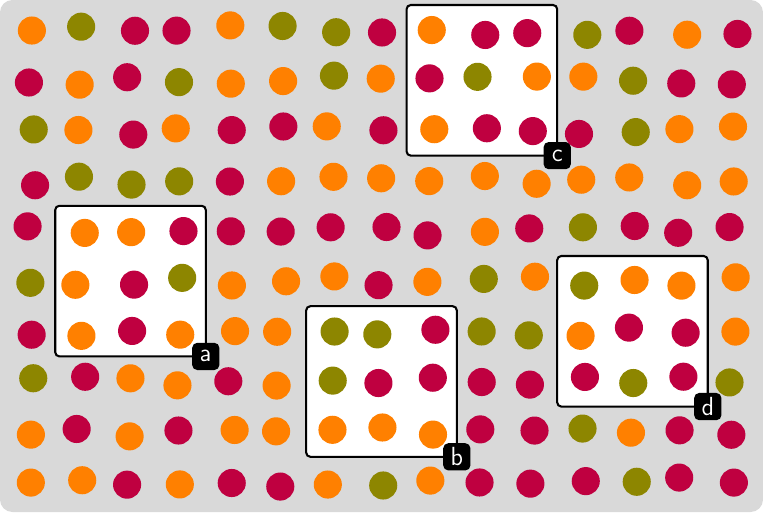}
 \caption{\small \textbf{Nuclei formation}
  translationally disordered state represented by 150 random samples from an \textcolor{olive}{\ch{A}}\textcolor{purple}{\ch{B2}}\textcolor{orange}{\ch{C2}} alloy (sample distribution \SI{19.3}{\%} \textcolor{olive}{A}, \SI{40.7}{\%} \textcolor{purple}{B}, and \SI{40.0}{\%} \textcolor{orange}{C}). The selected example nuclei show diverging compositions for the four small samples
  \textbf{a} \textcolor{olive}{11.1}|\textcolor{purple}{33.3}|\textcolor{orange}{55.5}\,\si{\%},
  \textbf{b} \textcolor{olive}{33.3}|\textcolor{purple}{33.3}|\textcolor{orange}{33.3}\,\si{\%},
  \textbf{c} \textcolor{olive}{11.1}|\textcolor{purple}{55.5}|\textcolor{orange}{33.3}\,\si{\%} and,
  \textbf{d} \textcolor{olive}{22.2}|\textcolor{purple}{44.4}|\textcolor{orange}{33.3}\,\si{\%}.
 }\,
 \label{fig:nck}
\end{figure}

{This important insight into solidification guides us in the design of the weighting approach for the database entries, by considering the earliest stage of nucleation as a statistical problem.}
Let us consider a small random alloy sample during its transition between the translationally disordered to the ordered state, as shown in Figure~\ref{fig:nck}.
Local compositions in the beginning phase deviate quite substantially from the overall composition.

\begin{figure*}[htb]
 \includegraphics[width=1.0\textwidth]{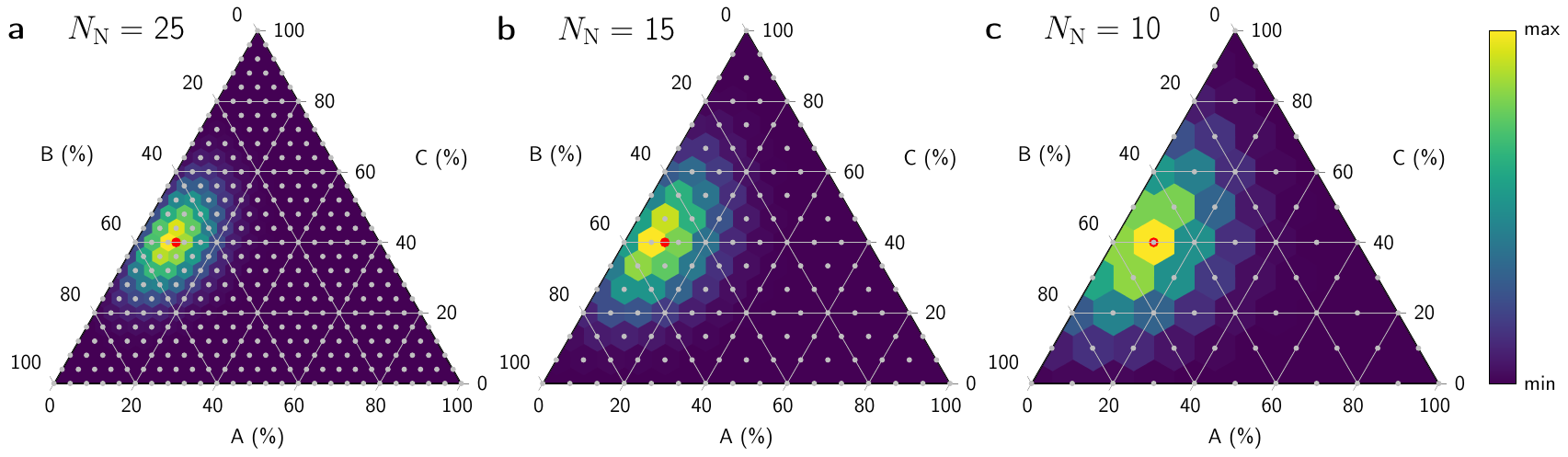}
 \caption{\small \textbf{Composition averaging} using the nuclei composition kernel. The red dot is placed at the nominal composition. Each grey point represents a possible composition given the number of atoms in the nuclei. The color around each point shows the relative weight of that region of the composition space. The composition area sharing the same weight increases with the decreasing number of atoms from (\textbf{a}) to (\textbf{c}).
 }
 \label{fig:comp}
\end{figure*}

To calculate the weight $w$ of an ordered sample, we use the multinomial discrete probability density function.
Here, the composition of a sample is distributed onto the chosen number of atoms in the nucleus $N_{\mathrm{N}}$.
So for each of the $N_\mathrm{S}$ species in the system, an integer number $N_{\mathrm{D}}$ is set by rounding the product from concentration fraction and $N_{\mathrm{N}}$.
As the sum of these integers can differ from $N_{\mathrm{N}}$, the discrepancy is corrected by adjusting the number with the largest error accordingly.
Then, the weight can then be expressed as:
\begin{align}
 N_{\mathrm{E}} &= \dfrac{N_{\mathrm{N}}!}{\prod_{j=1}^{N_\mathrm{S}}{N_{\mathrm{D},j}!}}\,,\\
 w &= N_{\mathrm{E}}\prod_{j=1}^{N_\mathrm{S}} p_j^{N_{\mathrm{D},j}}\,,
\end{align}
where $N_{\mathrm{E}}$ represents the number of equal outcomes and $p_j$ is the concentration of the $j$-th species in the translationally disordered phase.
As the number of atoms in the nucleus grows, the accessible compositions increase.
This leads to a distribution {centered} around the target composition, lowering the probability of deviation from the target.
In the limit of $N_{\mathrm{N}}\rightarrow\infty$ the distribution will resemble a Gaussian.
An example for an alloy with three species and different nuclei sizes is shown in Figure~\ref{fig:comp}.

A further desirable characteristic of the NCK is that the original composition is recovered even when the samples are randomly distributed in the composition space.
While this property is already helpful in {effectively averaging} database information for specific compositions, in the case of general material databases, an additional {step} is needed to avoid a shift in {the resulting overall} composition from highly sampled points.
{
For example, atomic compositions numbers having the largest ``greatest common factor'' have the smallest basis in the primitive cell, thus requiring less computational resources. As such, they tend to have significantly more calculations available in databases compared to other compositions.}
For this reason, all samples that belong to the same nucleus (gray points in Figure~\ref{fig:comp}) are binned together.
The combined entries created with this methodology are available in the Supplementary Datasets 2 and 3.

\section*{Results}
\label{sec:results}
\noindent{{\bf Constructing and integrating the thin-film dataset with the \AFLOWorg\ database } \\
  This comprehensive dataset was constructed by combining our labeled (amorphous, crystalline) experimental data of thin-film alloys produced by DC Magnetron co-sputtering~\cite{ding2014combinatorial} with theoretical properties of crystalline materials with similar compositions retrieved from the \AFLOWorg\ database.
The resulting dataset provides a unique platform to test if Soliquidy can be an integral part of a classifier capable of differentiating between the amorphous and crystalline labeled entries.
For further details on the \AFLOWorg\ computational database and the experimental setup employed in this study, readers are referred to the Method section.}
\\

\begin{figure}[ht!]
     \includegraphics[width=0.95\linewidth]{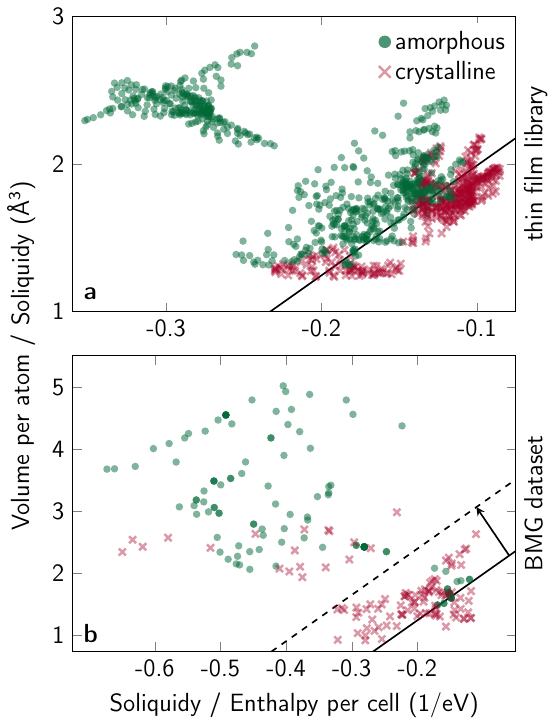}
 \caption{\small \textbf{Categorizing datasets.} (\textbf{a}) The solid black line separates the amorphous and crystalline systems in our thin-film library at a positive categorization rate of \SI{83.8}{\%}. In the case of (\textbf{b}) the public bulk metallic glasses (BMG) dataset, {the original separator needs to be shifted in parallel to accomplish the optimal} positive categorization rate of \SI{78.9}{\%} (dashed line).
 }
 \label{fig:comp_datasets}
\end{figure}

\noindent{{\bf \SISSO\ analysis } \\
  By exploring {the combined} {thin-film dataset} {and \AFLOWorg\ database} with \SISSO, we found several possible combinations of features able to tackle glass formation.}
The ten {best-performing classifiers} all rely on Soliquidy as part of the model {(quality is defined from the distances of wrongly classified entries from the separator boundary).
The best classifying manifold we found is the following:}
\begin{align}
 \dfrac{\text{volume per atom}}{\text{Soliquidy}} &= A \cdot \dfrac{\text{Soliquidy}}{\text{enthalpy per cell}} + B.
 \label{2dmanyfold}
\end{align}
This {two-dimensional locus combines: Soliquidy, volume per atom, enthalpy per cell, and two coefficients, $A=\SI{7.44}{\angstrom^3\eV}$ and $B=\SI{2.73}{\angstrom^3}$.}
By using Equation (\ref{2dmanyfold}), our experimental dataset can be divided by a simple linear equation into {the two categories amorphous and crystalline} (Figure~\ref{fig:comp_datasets}\textbf{a}), with a categorization success of \SI{83.8}{\%}.
This split means that out of 1126 samples, 943 are correctly categorized.
When leaving parts of the dataset out of the fit in a 5-fold cross-validation utilizing Support Vector Classification, the average positive classification rate is \SI{80.0}{\%}.
The isolated island in the upper left corner of Figure~\ref{fig:comp_datasets}\textbf{a} is formed by the entries from \ch{CuMgY} and \ch{AlCuZr}, two alloy systems known as good glass formers, that exhibit amorphous structures over the full experimentally covered concentration range.
{
  Overall, our positive categorization rate is similar to other approaches.
  For example, by using a general-purpose machine learning framework, Ward et al. obtained a characterization rate between \si{80-90}{\%}~\cite{wardIncludingCrystalStructure2017}.
  Other models, leveraging only composition information, can reach accuracies of 89\%~\cite{liEncodingAtomicStructure2022}.
  Our model is based on the glass-agnostic assumption -- the purposeful exclusion of features that focus on closeness to known glass formers.
  Training on such features would drive models to learn that compositional neighbors of glass formers are also quite likely glass formers
  without attempting to elucidate the underlying mechanism of solidification.
  Our method aims to predict --- with a certain degree of accuracy --- if an alloy at a specific composition can form an amorphous phase
  purely on the first-principle calculation of geometric and energetic features of crystalline structures.
}

\noindent{\bf Testing Soliquity on a \BMG\ public dataset } \\
To test the proposed classifier, we are also applying it to the public dataset
published by Ward in 2018~\cite{wardMachineLearningApproach2018,wardBulkMetallicGlasses2018}.
It needs to be noted that the published dataset contained over 150 entries with stoichiometries that did not correspond to the published works from where they were taken.
These mistakes were corrected before its use.

In Figure~\ref{fig:comp_datasets}\textbf{b}, all ternary compositions that are labeled in the public dataset as \textit{\BMGs}/\textit{none} are plotted.
As the classifier depends on \AFLOWorg\ data, only ternary alloy systems with at least 300 entries were considered.
The solid black line in Figure~\ref{fig:comp_datasets}\textbf{b} represents the same separator used for our experimental data.
It is clear that the separation would not be sufficient.
This is not very surprising as the experimental methods used in both cases are quite different.
The cooling rate of the samples in the thin-film library (\SI{\approx E8}{\kelvin\per\second} \cite{bordeenithikasemDeterminationCriticalCooling2017}) is orders of magnitude faster than the bulk cooling rates experienced by those in the public dataset.
It is to be expected that the higher effective cooling rate during the sputtering process led to a higher {ratio} of amorphous systems.
{Splitting the public dataset optimal reveals that only a parallel shift of the original separator into the amorphous region was necessary.
By changing the $B$ parameter of the model from \num{2.73} to \SI{3.89}{\angstrom^3} (dashed line in Figure~\ref{fig:comp_datasets}\textbf{b}) we can achieve a positive categorization rate of \SI{78.9}{\%}.}
The 5-fold cross-validation for this dataset results in a positive classification rate of \SI{73.0}{\%}
While a considerable amount of sample points are misclassified, the trend inside the corresponding alloy systems still follows the same separation.
Overall, we can observe a change in the balance between the samples' energetically driven crystallization and the geometric hurdles opposing it.
The shift demonstrates that energetically less favorable structures are more important at lower cooling rates.

\begin{figure}
 \includegraphics[width=0.8\linewidth]{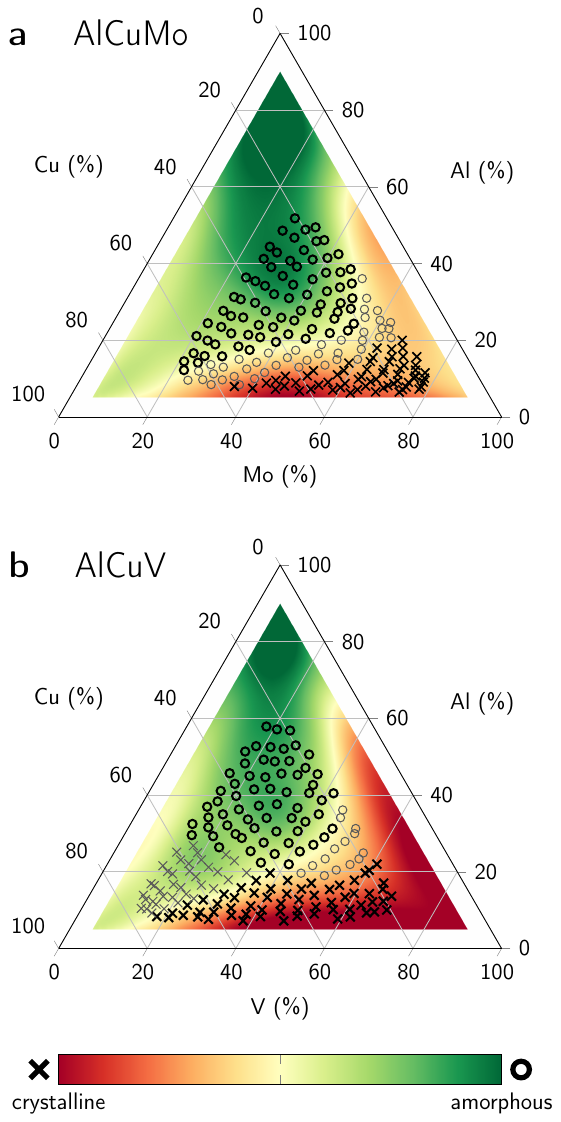}
 \caption{\small \textbf{Separation distance} for the training set alloys (\textbf{a}) \ch{AlCuV} and (\textbf{b}) \ch{AlCuMo} are shown as the colored background.
 The symbols on top indicate the experimental outcome in the thin-film library.
 Bold black symbols show regions with correct predictions.}
 \label{fig:comp_experiment}
\end{figure}

In addition to a binary separation, we can also consider the distance from the dividing line to visualize the distribution within the ternary alloy system.
We define the separation distance $d_{\mathrm{S}}$ as:
\begin{align}
 d_{\mathrm{S}} &\equiv \dfrac{B \cdot \dfrac{\text{volume per atom}}{\text{Soliquidy}}-A \cdot \dfrac{\text{Soliquidy}}{\text{enthalpy per cell}}}{\sqrt{A^2+1}}.
\end{align}
Positive/negative values of $d_{\mathrm{S}}$ indicate an amorphous/crystalline  structure.
In Figure~\ref{fig:comp_experiment} this is plotted for \ch{AlCuV} and \ch{AlCuMo} respectively.

The results demonstrate a strong correlation between the newly introduced {classifier utilizing Soliquidy} and the tendency of a specific composition to form an amorphous solid phase.
By using \SISSO, we generated a straightforward relationship between the information available in the \AFLOWorg\ database and our experimental data.
The transferability of the trends observed in our data to a public dataset reinforces our confidence in the found {classifier and the Soliquidy descriptor itself}.
The shift of {one} coefficient for the separation line in the case of the public BMG datasets is expected, as the coefficients capture the environment of the observed phase transition.
In this case, the cooling rates in our experimental setup are significantly higher than for the public dataset.
Overall, these findings validate Soliquidy's capability to capture relevant geometrical information to investigate phase transitions.

The reliance on a consistently filled material database limits the number of systems that could be examined from the public dataset, as we only considered alloys with at least 300 entries.
To further improve the comparability between systems and reduce the impact of over-sampled compositions, a subset of prototypes should be selected and populated for all systems in question.
This approach leads to a selection process in well-explored systems, while in under-explored systems, it provides a roadmap of required calculations.
For this study, we only employed the prototype approach to expand the number of entries in alloy systems, which had experimental results but needed to be computationally explored more thoroughly.

{
In this article we have introduced Soliquity, a descriptor which measures confusion in an atomic geometrical organization. Soliquity is calculated with the Euclidean transport cost of each atom moving through the surrounding Voronoi cell to its final equilibrium position. By extending our library of thin-film alloys and combining it with geometrical/energetic quantities extracted from the \AFLOWorg\ computational database, we show that the combination of Soliquity and formation enthalpies can generate effective classifiers for metallic-glass formation. Most importantly, the glass-agnostic assumption does not require any knowledge of glass formation of nearby compositions.  Soliquity can then be used to tackle other kinetic-controlled transformations where translational disorder plays a critical role, such as spinodal decomposition or solidification.
}

\section*{Methods}
\noindent {\bf \texttt{AFLOW.org} database } \\
To enable predictions regarding the glass forming abilities of metallic systems, a reliable data source is required.
For this work, we use entries from the \AFLOWorg\ database~\cite{curtarolo:art190}.
The systems collected are computed based on the \AFLOW\ standard~\cite{curtarolo:art104}, enabling a consistent baseline for comparisons and analytical tasks.
With over 4 million entries, many alloy systems can be explored directly.
The inclusion of ICSD structures ensures that relevant and representative structures are available in different alloy systems.
Overall, \num{9805} entries are used here. The list, grouped by alloy system, is available in Supplementary Dataset 1.
In cases of under-explored systems, the \AFLOW\ library of
crystallographic
prototypes~\cite{curtarolo:art121,curtarolo:art145,curtarolo:art173,eckertAFLOWLibraryCrystallographic2024} can be used to fill in the gaps in a reproducible way.

\noindent {\bf SISSO (Sure Independence Screening and Sparsifying Operator) } \\
The algorithm combines symbolic regression and compressed sensing~\cite{SISSO} to identify mathematical functions that best predict a target property of a dataset.
It can model complex phenomena using simple descriptors for regression and classification tasks from tens to thousands of data points.
For this work, we are utilizing the \SISSO++ package~\cite{purcellSISSOImplementationSureIndependenceScreening2022}, an open-source implementation of the \SISSO\ method.

Data points of ternary alloys from the \AFLOWorg\ database are combined, using the nuclei composition kernel, into a collection of features for each composition available in the experimental dataset.
For this, a nucleus size of $N_{\mathrm{N}}=25$ is selected, ensuring that the nucleus is significantly smaller than the critical size while still focusing on \AFLOWorg\ entries close to the experimental composition.
Overall, a collection of \num{1126} entries, matching the experimental data across six ternary metal alloy systems (\ch{AlCuMo}, \ch{AlCuV}, \ch{AlCuZr}, \ch{AlMoPd}, \ch{AlMoV}, \ch{CuMgY}) was created.
Each entry is labeled either amorphous or crystalline, based on the experimental results.
Both labels serve as the target for the training.

After a first exploration using all the available properties in the database, we removed the ones having little or no correlation, and we focus on the following feature subset: energy per atom (\si{\eV}), enthalpy per atom (\si{\eV}), enthalpy per cell (\si{\eV}), formation enthalpy per atom (\si{\eV}), Soliquidy (unitless), and volume per atom (\si{\angstrom^3}).
All features are based on computed quantities.
To allow \SISSO\ to be effective and efficient, the exploration was limited to the following set of mathematical operations: $A+B$, $A-B$, $|A-B|$, $A\cdot B$, $A/B$, $|A|$, $\exp(A)$, $\exp(-A)$, $\ln(A)$, $A^{-1}$, $A^2$, $A^3$, $A^6$, $\sqrt{A}$, and $\sqrt[3]{A}$.

\noindent {\bf Experimental {thin-film library} preparation } \\
The thin-film library is fabricated through confocal DC Magnetron co-sputtering (AJA International ATC2200).
Sputtering targets of purity better than \SI{99.95}{\%} are used (AJA International and Kurt J. Lesker Company).
Silicon wafers of \SI{4}{\inch} with a \SI{100}{nm} thick thermal oxide layer are used as substrates.
Prior to sputtering, the chamber is evacuated to a pressure of \SI{1E-06}{\torr} or better.
The films are sputtered in ultra-high purity argon gas at a pressure of \SI{5.8E-03}{\torr}.
Sputtering guns are arranged in a tetrahedral geometry with the substrate.
This arrangement leads to a compositionally graded film, where the film at each point on the library represents an alloy with a different chemistry.
Typically, the species atomic fraction can vary by up to \SI{40}{\%} within an approximate range of 10 to \SI{95}{\%}.
If one considers that a distinction of alloys require at least one atomic percent \cite{Li_acscombsci_numMG_2017}, up to 1000 alloys
can be represented within one compositional library~\cite{ding2014combinatorial, kubeMetastabilityHighEntropy2020, liDatadrivenDiscoveryUniversal2022, kubeCompositionalDependenceFragility2022}.

The chemical composition of the various alloys in the thin-film library is characterized using high-throughput energy-dispersive x-ray spectroscopy (Helios G4 Focused Ion Beam – Scanning Electron Microscopy with UltraDry EDX detector, \SI{25}{\kilo\V} accelerating voltage).
The atomic structure is characterized using high-throughput x-ray diffraction (XRD), performed at beamline 1-5 at the Stanford Synchrotron Radiation Lightsource (SSRL) at SLAC National Accelerator Laboratory (\SI{12.7}{\kilo\eV} photon energy, automated xy-stage to perform measurement across a \SI{6}{\milli\meter} square grid)~\cite{kubePhaseSelectionMotifs2019}.
Each ternary alloy system is measured through 177 evenly spaced $xy$ positions representing 177 alloys with unique compositions.
{The only exception is \ch{CuMgY}, measured on a finer grid containing 241 points.}

\vspace{2mm}

\section*{Declaration of Competing Interest}
The authors declare that they have no competing financial interests
or personal relationships that could have appeared to influence the work
reported in this article.

\section*{Acknowledgments}
The authors thank Marco Esters, Scott Thiel, Andriy Smolyanyuk, Corey Oses, Ohad Levy, Arrigo Calzolari, Yoav Lederer, Gianluca Di Muro, and Michael Mehl for fruitful discussions.
This research was supported by the Office of Naval Research under grant N00014-20-1-2200 and N00014-20-1-2225.
This work was supported by high-performance computer time and resources from the DoD High Performance Computing Modernization Program (Frontier).
We acknowledge Auro Scientific, LLC for computational support.

\section*{Author Contributions}
H.E.\ and S.C.\ conceived the research and designed the theoretical framework.
H.E., A.G.\ and A.C.Z.\ prepared, performed and processed the calculations.
S.A.K.\ and J.S.\ synthesized and measure the thin-film library.
H.E., S.D., X.C.\ and S.C.\ wrote the manuscript with contributions from J.S., S.A.K., D.H., N.H., S.D.G., S.Z.\ and A.v.d.W.
All authors have given approval to the final version of the manuscript.

\section*{Data availability}
The raw data employed to generate the predictions are freely available from the \AFLOWorg\ database.
The list of utilized \AFLOWorg\ entries and the processed data are included in the Supplementary Information as Supplementary Datasets 1-3.

\section*{Code availability}
\AFLOW4 was used in this study and its source code can be accessed under GPL via this link \url{https://aflow.org/install-aflow/}.

\newcommand{\Ozolins}{Ozoli{\c{n}}{\v{s}}}

\end{document}